\documentclass[aps,groupedaddress,showpacs,twocolumn]{revtex4}

\newif\ifpdf
\ifx\pdfoutput\undefined
\pdffalse 
\else
\pdfoutput=1 
\pdftrue
\fi

\ifpdf
\usepackage[pdftex]{graphicx}
\else
\usepackage{graphicx}
\fi

\begin{document}
\ifpdf
\DeclareGraphicsExtensions{.pdf, .jpg}
\else
\DeclareGraphicsExtensions{.eps, .jpg}
\fi
\def\grad{\vec{\nabla}}
\def\div{\vec{\nabla}\cdot}
\def\curl{\vec{\nabla}\times}
\def\DDt{\frac{d}{dt}}
\def\ddt{\frac{\partial}{\partial t}}
\def\ddx{\frac{\partial}{\partial x}}
\def\ddy{\frac{\partial}{\partial y}}
\def\lap{\nabla^{2}}
\def\divv{\vec{\nabla}\cdot\vec{v}}
\def\gradS{\vec{\nabla}S}
\def\vvec{\vec{v}}
\def\wc{\omega_{c}}
\def\<{\langle}
\def\>{\rangle}
\def\Tr{{\rm Tr}}
\def\Csch{{\rm csch}}
\def\Coth{{\rm coth}}
\def\Tanh{{\rm tanh}}
\preprint{Submitted to Phys. Rev. Lett., Jan 11, 2001}

\title{Quantum Liouville-space trajectories for dissipative systems}

\author{Jeremy B. Maddox}\email[email:]{jmaddox@bayou.uh.edu}
\author{Eric R. Bittner}
\email[email:]{bittner@uh.edu}
\affiliation{Department of Chemistry, University of Houston,Houston 
Texas 77204}


\date{\today}

\begin{abstract}
In this paper we present a new quantum-trajectory based
treatment of quantum dynamics suitable for dissipative systems. 
Starting from a de Broglie/Bohm-like representation of the 
quantum density matrix, we derive and define quantum equations-of-motion 
for Liouville-space trajectories for a generalized system coupled to 
a dissipative environment.
Our theory includes a vector potential which mixes forward and backwards
propagating components and non-local  
quantum potential which continuously produces coherences in the system.
These  trajectories are then used to propagate an adaptive Lagrangian grid
which carries the density matrix, $\rho(x,y)$, 
and the action, $A(x,y)$,
thereby providing a complete 
hydrodynamic-like description of the dynamics. 
\end{abstract}

\maketitle


The causal or hydrodynamic trajectory 
model of quantum mechanics provides a
useful glimpse into the dynamics underlying the 
quantum wavefunction.\cite{QTM}
Starting from the time-dependent
Schr\"odinger equation, one can derive quantum trajectory
equations by casting the wavefunction in polar form, $\psi
=R\exp(iS/\hbar)$ and subsequently separating 
$\dot\psi$ into real and imaginary
components.
The resulting equations are a continuity equation for the
quantum density, $\rho=R^{2}$, and ``Newtonian'' equations for
trajectories.
The distinguishing feature of 
these quantum equations-of-motion
is the presence of the 
non-local quantum potential, $Q$, which
both introduces non-local coupling between the
trajectories and  
represents a differential geometric constraint between the extrinsic
curvature invariants of a surface generated by $z = C \ln R$ and the
action per unit volume of a trajectory element.   
This prescription introduced independently by de Broglie\cite{Broglie26}
and Madelung\cite{Madelung26},
and later developed by Bohm\cite{Bohm52} has been 
used primarily as an interpretive tool where one
first solves the time-dependent
Schr\"odinger equation for the wavefunction, $\psi(x,t)$,
and then  uses $\psi$
as an ancillary field to drive an ensemble of quantum
trajectories.  
From this construction one can develop 
interpretive models of barrier tunneling, arrival times, and 
various other quantum effects based upon these trajectories.~\cite{QTM}
Furthermore, there has been a recent surge of activity to
use this approach to develop new adaptive-grid based  methods
for performing quantum dynamical calculations. Such {\em ab initio}
approaches construct $\psi$ directly from the dynamical information 
obtained in computing the trajectories.
\cite{ref1,ref2,ref3,ref4,ref5} 
While it is unclear at this time
whether such approaches offer considerable computational advantages
over more standard finite-basis set or fixed grid approaches,
the appeal is that by moving to a
particle-based description, computational effort 
required scales almost linearly with the number of 
trajectories introduced.

In this letter we derive a 
causial-trajectory approach 
for the quantum density matrix under
dissipative conditions.  So far as we know, this is the
first presentation of such an approach and certainly is
the first {\em ab initio} application. The 
formal construction presented here is similar in spirit to the 
Wigner phase-space representation which is widely used
for quantum dissipative dynamics in both the quantum and semi-classical
regimes; however, our approach does not require the definition of 
a quantum phase space and  
retains all of the physical properties associated 
with the quantum density matrix. Secondly, we describe an 
{\em ab initio} approach for computing the trajectories based upon 
a finite-element/least squares proceedure.  
This provides a highly efficient and adaptive scheme 
which can be applied to a wide variety of systems. Moreover, the 
computational effort 
scales almost linearly with number of finite-elements/trajectories.
In this letter  we focus upon 
the construction of the underlying theory 
with application towards the relaxation of
a tagged oscillator in contact with an environment.

The theory is initiated by writing the density matrix as $\rho(x,y) =
\psi(x)\psi^{*}(y)$.  If $\rho$ represents the entire system+bath 
ensemble, its evolution is given by the Liouville-von Neumann equation,
  $\dot{\rho}= L\rho$,
where $L$ is the full Liouville operator which we decompose into system and 
bath components: $L = L_{s} + L_{b} + L_{sb}$. If we take the bath as 
an ensemble of harmonic oscillators with masses, $m_n$, and frequencies, $\omega_n$,
which is at thermal equilibrium at $t=0$ coupled to the system 
coordinate, $x$, via a linear
combination $x = \sum_n c_n q_n$, then
the response of the bath to this coupling is 
encoded in the correlation functions
$\hbar(\nu(t) + i \eta(t)) = \langle x(t) x(0)\rangle/_T$ where 
$\langle\ldots\rangle_T$ denotes the expectation value taken 
in the thermal equilibrium state of the bath. 
The real and imaginary parts of this
correlation function are the noise and dissipation kernels. 
This procedure is starting 
point of many quantum statistical mechanical treatments and the result 
is either a master equation for $\dot{\rho}_{s}$, or a path-integral 
treatment for propagator. Various forms of the master equation  
have been presented in the literature by a number of groups including 
Caldeira and Leggett\cite{CL}, 
Unruh and Zurek\cite{UZ}, and by Hu and
co-workers\cite{Hu}.  Each of these is slightly different in final form due
to the various approximations made over the course of the derivation.
Consequently, each posses certain regions of validity. 
Without loss of generality, we will focus our attention here on the
Caldeira-Leggett\cite{CL} (CL) model whereby the noise and dissipation terms
take the form $\nu(t) = {\gamma \hbar}/{\lambda^2}\delta(t)$ and 
$\eta(t) = - \gamma m\delta'(t)$. This model is strictly valid 
in the high-temperature regime 
where the thermal excitations in the bath dominate and fluctuations 
of the vacuum can be ignored.  Here, the master equation for the reduced density matrix is given by
\begin{eqnarray}
\partial_t \rho =\tilde{L}_s\rho
- \gamma (x - y)(\partial_x \rho - \partial_y \rho) 
- \frac{\gamma}{2\lambda^2}(x - y)^2\rho.
\end{eqnarray}
Here, $\tilde{L}_s$ denotes the system Liouvillian
where the potential energy term has been renormalized due 
to the interaction with the bath. 
For a harmonic system, the renormalized  oscillator 
frequency is related to the
bare frequency,  $\omega$, by
 $\tilde\omega^2 = \omega^2-2\gamma\omega/\pi$.
The last two terms are related to interaction between the 
system and the bath. The second being responsible 
for population relaxation and the third for coherence 
relaxation.  
Finally $\lambda = \hbar/\sqrt{2mkT}$ is the thermal de Broglie wavelength 
and $\gamma$ is the coupling strength.

To make the connection to a quantum trajectory scheme,
we write the density matrix in complex 
polar form,
$\rho_{s}(x,y)=\exp(g(x,y) + i A(x,y)/\hbar)$,
 and insert this into the master equation for
$\dot{\rho}_{s}(x,y)$.  In what follows, we shall take $x$ as the 
coordinate associated with $\psi(x)$ and $y$ as the coordinate 
associated with the complex conjugate, $\psi^{*}(y)$. Our Liouville 
space is then the two-dimensional configuration space $\{x,y\}$ 
with phase-space defined as $\{x,p_x,y,p_y\}$ 
where $p_x = \partial_xA$ and $p_y = -\partial_yA$. 
The $-$ sign in the definition of the canonical momentum in $y$ 
reflects the time-reversed dynamics in the $y$ direction. 
In a geometric-optical construction, the wavefront for the density 
matrix is given by contours of constant 
$S(x)+S(y) = c$ with velocities and momenta are both 
normal to these curves.
For non-dissipative dynamics, 
the Hamilton-Jacobi equation for the action
$A(x,y)$ is separable into forward (in $x$)
and backward (in $y$) propagating components:
\begin{eqnarray}
\frac{\partial A}{\partial t} 
=-\frac{1}{2m} \left(
p_x^2 -p_y^2\right)
&-& {\cal V} - {\cal Q},
\end{eqnarray}
where ${\cal V} = V(x)-V(y)$ and ${\cal Q} = Q(x)-Q(y)$.
The quantum potential,
$Q(x) = - \hbar^2(\partial_{x}^{2}g + \partial_{x}g\partial_{x}g)/2m$,
introduces non-local interactions between the trajectories
but does not couple motion in $x$ and $y$. 
The resulting equations of motion take 
the form, $mv_{\mu} = p_{\mu}$, where upon taking the time derivative
yields Bohm's quantum equations of motion in $x$ and $y$:
$m\ddot{x}_\mu = \pm\partial_\mu (V + Q)$ ($-$ for $x$ and $+$ for $y$).
We also have the continuity equation for $g$:
 $2\dot{g}=-\partial_\mu v^\mu$.

\begin{figure}[ht]
\includegraphics[width=3.0 in]{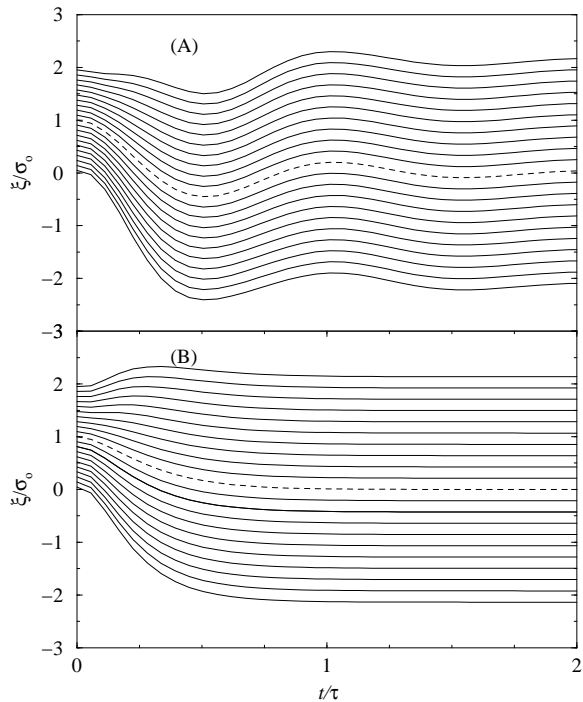}
\caption{
Diagonal trajectories for underdamped (A) and critically damped (B) cases
at $kT = 2.5\hbar\omega$. Dashed line corresponds to a semi-classical approximation for  $\<\xi(t)\>$ (c.f.Eq.~\ref{eom-xi})}
\label{fig1}
\end{figure}

Taking the CL equation and 
using the substitution $\log \rho = g + iA/\hbar$ as before 
again yields a Hamilton-Jacobi equation for $A(x,y)$:
\begin{eqnarray}
\partial_t A &+& 
\frac{1}{2m}(p_x^2 - p_y^2)-\gamma(x-y)(p_x - p_y)
+ {\cal V} + {\cal Q} = 0.\nonumber
\end{eqnarray}
However, this equation now sports an explicit coupling between the
forward and backward propagating paths. 
This dissipative coupling term involving $\gamma(x-y)$ 
can be cast as a
vector potential,$\Gamma_\mu = \pm \gamma(x-y)$ where 
$-$ is for $x$ and $+$ is for $y$. 
Using this we define the components of the {\em material
velocities} as $m\dot{x}_{\mu} = p_\mu + \Gamma_{\mu}$ which 
describe the velocities of the particles in the presence of dissipation.

At this point we find  
it easier to work in the rotated coordinate frame given by 
$\xi = (x+y)/\sqrt{2}$ 
and $\eta = (y-x)/\sqrt{2}$. In this frame, the quantum Lagrangian is 
\begin{eqnarray}
{\cal L}
=-m \dot{\xi}\dot{\eta} + 2 m \gamma \eta\dot{\xi}-\cal{Q} - {\cal V}.
\end{eqnarray}
Here the quantum potential is given by
\begin{eqnarray} 
{\cal Q} = \frac{\hbar^2}{m}(\partial_\xi\partial_\eta g 
+ \partial_\xi g \partial_\eta g),
\end{eqnarray}
 and ${\cal V} = V(x)-V(y)$.
In this new frame, the canonical momenta are given by 
$p_\xi = -\partial {\cal L} /\partial\dot{\eta}$ and 
$p_\eta = -\partial {\cal L}/\partial\dot{\xi}$.
This modified relation between the canonical momenta and the 
Lagrangian
is again due to the time-reversed dynamics in $y$. 
The resulting material velocities are then 
\begin{eqnarray}
\dot{\eta} = p_\eta/m + 2 \gamma  \eta
\,\,{\rm and}\,\,
\dot{\xi} =p_\xi/m.
\end{eqnarray}
Finally, the continuity equation for  $g$ is given 
in the rotated frame as 
\begin{eqnarray}
\frac{dg}{dt} = -\frac{1}{2}
\left[ \partial_\xi\dot{\xi} + \partial_\eta\dot{\eta}
\right] + \gamma - \frac{2\gamma}{\lambda^2}\eta^2.
\end{eqnarray}
This  equation 
includes a term proportional to
$\eta^2$ which dampens components of the
density matrix which are far from the diagonal axis
leaving the
elements  along the $\xi$-axis unchanged. 
This is the process of decoherence, i.e. 
the dynamical diagonalization of the
density matrix due to the interaction with the environment.
\cite{footnote}

We take as an example problem the case of harmonic oscillator embedded
in a dissipative bath. Such a system could easily represent a
vibrational coordinate of a molecule that has been displaced from its
equilibrium position by some action at $t=0$. 
Our computational methodology.
is based upon the moving weighted
least-squares (MWLS) method reported by Lopreore and Wyatt
and reader is referred to Ref.\cite{ref1} 
for details of this approach and
to Ref. \cite{ref2} for our
implementation as applied to quantum wavepacket dynamics.
\cite{code}

Taking $k$ as the renormalized
force constant and $\omega^2 =k/m$, 
the quantum Lagrangian is
\begin{eqnarray}
{\cal L} = - m\dot{\xi}\dot{\eta}
+ 2m\gamma\dot{\xi}\eta 
+ {k} \xi\eta 
- Q.
\end{eqnarray}
Using the Euler-Lagrange equations, we can derive equations of motion 
for the Liouville space quantum trajectories: 
\begin{eqnarray}
\ddot{\xi} &=& -\omega^2\xi 
- \frac{1}{m}Q_\eta-2\gamma\dot{\xi}\label{eom-xi}\\
\ddot{\eta} &=& -\omega^2\eta 
- \frac{1}{m}Q_\xi+2\gamma\dot{\eta}.\label{eom-eta}
\end{eqnarray}
Defining the semi-classical limit as $\hbar\rightarrow 0$ so that $Q$
vanishes, the equations of motion are trivial to solve.  
For motion in $\xi$ we
have damped harmonic motion which is easy to understand in terms 
of pure populational relaxation. 
However,  in $\eta$, the trajectories diverge
towards $\pm\infty$. 

\begin{figure}[hb]
\includegraphics[width=3.0 in]{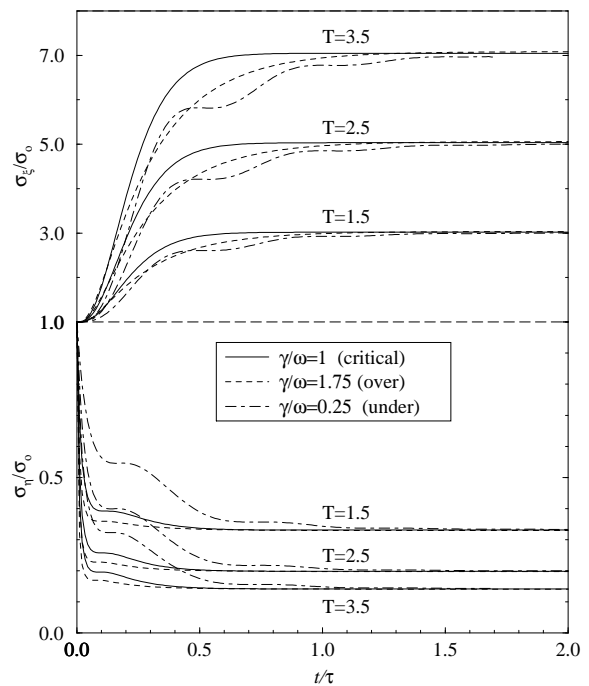}
\caption{Relaxation of $\sigma_\xi$ and $\sigma_\eta$ at various
temperatures in the overdamped $(\gamma/\omega = 1.75)$,
 underdamped $(\gamma/\omega = 0.25)$, and critically damped
 $(\gamma/\omega = 1)$ 
regimes.}
\label{fig2}
\end{figure}

This introduces a small problem in our computational scheme, 
namely, all of the non-diagonal grid points are swept away
as the calculation progresses.
To circumvent this, we {\em re-meshed}
the grid periodically by discarding the off-diagonal
trajectories and 
re-initiating $\rho$, $g$, and $A$ at the 
new points via least-squares interpolation.  We also took advantage of
the parity of these functions under $\eta\rightarrow -\eta$
by explicitly propagating only the upper triangle of each 
and selecting polynomial bases which carried the appropriate 
symmetry.  To whit:  
$g(\xi,\eta)=g(\xi,-\eta)$, 
$A(\xi,\eta)=-A(\xi,-\eta)$,
$Q(\xi,\eta)=-Q(\xi,-\eta)$,
$v_\xi(\xi,\eta) =v_\xi(\xi,-\eta)$,
$v_\eta(\xi,\eta) =-v_\eta(\xi,-\eta)$.
These symmetries were enforced throughout the calculation.

As an initial example, we consider the relaxation of 
a density matrix corresponding to the 
ground state displaced from equilibrium by a shift in $x$. 
\begin{eqnarray}
g(0)= 
g_o - \frac{1}{2\sigma_o^2}((\xi-\xi_o)^2
+ \eta^2)
\end{eqnarray}
In the plots shown here, we scaled the time units in terms of 
the oscillator period $\tau = 2\pi/\omega$ and lengths by the initial 
width of the gaussian density matrix, $\sigma_o^2 = \hbar/m\omega$.

In the Fig. 1, we show an ensemble of quantum trajectories corresponding 
to diagonal element of the density matrix (i.e. $\eta = 0$) at
$kT = 2.5\hbar\omega$ in the underdamped $(\gamma/\omega = 0.25)$
 and critically damped $(\gamma/\omega = 1)$ regimes.  The superimposed 
dashed line is the semi-classical trajectory obtained by solving
Eq.\ref{eom-xi} in the limit of $\hbar\rightarrow 0$.  As one expects, the
the semi-classical result follows the peak of the gaussian. 
Furthermore, the diagonal trajectories are not allowed to cross one another. This is a directly analogous to the non-crossing rule for ordinary
quantum trajectories.   Each diagonal element carries a unique 
population trajectory which must remain single-valued. As time progresses
the trajectories become stationary corresponding to thermal equilibrium.

In Fig. 2 we show the time evolution of the gaussian coefficients
$\sigma_\xi$ and $\sigma_\eta$ for various temperatures and 
coupling constants.  
At thermal equilibrium, these quantities should become
stationary and  
can be computed from equilibrium statistical mechanics,
\begin{eqnarray}
\sigma_{\xi,eq}^2 = \sigma_o^2\Coth(\hbar\omega\beta/2) \\
\sigma_{\eta,eq}^2=\sigma_o^2\Tanh(\hbar\omega\beta/2).
\end{eqnarray}
For the critically damped case, we see 
exponential relaxation back to equilibrium.  For $\sigma_\xi$, 
the relaxation is largely independent of temperature 
and each case reaches the equilibrium value at $t/\tau = 1$. 
The relaxation of the coherence width, $\sigma_\eta$, shows a 
strong temperature dependence, relaxing faster at higher temperatures.
Moreover, the time-scale for coherence 
relaxation is considerably shorter than 
in $\sigma_\xi$. 
This is due to the fact that
the thermal de Broglie wavelength, $\lambda$, sets both a length
and time-scale for decoherence.\cite{zurek} As temperature increases, this 
length decreases and the relaxation rate increases. 
The underdamped case shows rather interesting dynamics by 
relaxing through a series of intermediate plateaus.  Finally, the
overdamped case relaxes slower than the critically damped without the 
plateaus seen in the underdamped case.  

The picture we are lead to is that the dissipative coupling
to the environment causes a net flux of trajectory elements
(representing population coherence information) towards
$\eta\rightarrow\pm\infty$ and causes the populations along the 
$\eta = 0$ axis to relax to some lowest energy configuration.
At $T=0$ this would be the quantum ground state of the system.
At finite temperature the system 
becomes stretched in $\xi$ reflecting a thermal distribution of 
energy states. Furthermore, the coherence length as set by 
the de Broglie wavelength,  becomes more and more short ranged
as $T$ increases
causing the system to become localized in $\eta$, 
effectively diagonalizing the density matrix.  
In the equations-of-motion, both the
quantum potential {\em and} the vector potential, $\Gamma$, 
force particles to 
stream outwards in $\eta$ away from the diagonal.
Consequently, even when the system becomes stationary 
(i.e. $\dot g = 0$) 
the trajectories themselves remain in constant motion reflecting the 
continuous influx and efflux of energy between the system and the bath
and the continuous streaming of coherence from the 
system into the bath degrees of freedom.  This continual ``production''
of coherence  is ultimately traced to the non-local nature of the  
quantum potential, ${\cal Q}$. 

We present here a novel extension of the de Brogile/Bohm
quantum theory of motion
into Liouville space and use this to propagate quantum trajectories for 
the density matrix of a system in contact with a thermal bath.  This 
approach opens a clear avenue to 
a number of new semi-classical and quantum-classical
approximations for the quantum density matrix. Furthermore, 
the formalism itself offers an interesting dynamical twist to interpreting
decoherence and population relaxation. 
 
This work was supported in part by the National Science Foundation, 
the Robert Welch Foundation, and by the State of Texas Advanced
Research Program. We thank Prof. R. E. Wyatt (U. Texas) for 
many discussions over the course of this work.

%




\begin{thebibliography}{10}
\bibitem{footnote}
The $+\gamma$ in this equation is countered by 
a $-\gamma$ obtained upon inserting $\dot\eta$ and 
does not lead to $g$ blowing up uniformly in time.
\bibitem{code}
A copy of our code is available from the authors upon request.
\end{thebibliography}
\end{document}